\documentclass[%
 reprint,
 amsmath,amssymb,
 aps, prc
]{revtex4-2}

\usepackage{graphicx}
\usepackage{dcolumn}
\usepackage{physics}
\usepackage{amsmath}
\usepackage{amssymb}
\usepackage{bm}
\usepackage{hyperref}
\hypersetup{
    colorlinks=true,
    linkcolor=blue,
    filecolor=blue,
    urlcolor=blue,
    citecolor=blue,
}

\begin{document}
\preprint{APS/123-QED}
\title{Nuclear Excitation by Inelastic Positron Scattering in Hydrogenlike  $^{229}\mathrm{Th}^{89+}$ Ions}

\author{Bo Liu}
\affiliation{School of Information and Electrical Engineering, Hangzhou City University, Hangzhou 310015, China}
\author{Wu Wang}
\email[Corresponding author: ]{wangwu@itp.ac.cn}
\affiliation{Institute of Theoretical Physics, Chinese Academy of Sciences, Beijing 100190, China}

\author{Yong Li} 
\email[Corresponding author: ]{yongli@hainanu.edu.cn}
\affiliation{Center for Theoretical Physics, School of Physics and Optoelectronic Engineering, Hainan University, Haikou 570228, China}

\begin{abstract}
We develop a relativistic theory of nuclear excitation by inelastic
positron scattering (NEIPS) with nuclear hyperfine mixing (NHM)
explicitly included and apply it to hydrogenlike
$^{229}\mathrm{Th}^{89+}$ ions. The target ion is described by dressed
hyperfine states, while the incident and outgoing positron continuum
states are treated within a relativistic distorted-wave approach. We
calculate hyperfine-resolved excitation cross sections for incident
positron energies from $10^{2}$ to $10^{5}~\mathrm{eV}$. The cross
sections increase strongly with the positron energy, with the
$F_g=2\rightarrow F_e=2$ channel exhibiting the largest value.
Comparison with NEIPS of the bare
$^{229}\mathrm{Th}^{90+}$ nucleus shows that NHM enhances the excitation
cross sections by about four orders of magnitude. These results demonstrate that NHM provides an efficient mechanism for enhancing NEIPS and broaden the range of nuclear excitation processes in which NHM plays a significant role.
\end{abstract}

\maketitle

\section{introduction}\label{sec:introduction}
The  \(^{229}\mathrm{Th}\) nucleus possesses an exceptionally low-lying isomeric state with an excitation energy of about \(8.4~\mathrm{eV}\) \cite{PhysRevLett.132.182501,PhysRevLett.133.013201,Zhang2024}. This unusual property makes \(^{229}\mathrm{Th}\) a particularly attractive system for investigating the interplay between nuclear and electronic degrees of freedom. A variety of processes involving electrons have therefore been explored for exciting or manipulating the isomeric state. These include electronic bridge \cite{Tkalya1992,PhysRevLett.105.182501, PhysRevLett.124.192502,MULLER201784,PhysRevC.100.044306,PhysRevLett.125.032501,PhysRevC.102.024604,Li2023} and hyperfine electronic bridge \cite{Wang2024-2,Wang2025_2,Liu2026Shortcut} processes,
laser-driven electron recollision
\cite{PhysRevLett.127.052501,PhysRevC.106.024606,
	PhysRevResearch.5.043232},
nuclear excitation by electron capture
\cite{Qi2023,Zhao2024,Xu2026NEEC},
and nuclear excitation by inelastic electron scattering (NEIES)
\cite{PhysRevLett.124.242501,PhysRevC.106.044604,Xu2024NEIES,Wang2026Cascade}. For more comprehensive discussions of \(^{229}\mathrm{Th}\), electron-nucleus interactions, and nuclear excitation mechanisms, we refer the reader to recent reviews~\cite{Lu2026Review,Xu2026Review,Xiao2026Review}.

Another important electron-nucleus coupling effect is nuclear hyperfine mixing (NHM), in which the hyperfine interaction admixes different nuclear states through their coupling to bound electrons \cite{Lyuboshitz_1966}. Early studies of NHM were largely restricted to $^{229}\mathrm{Th}$  \cite{Wycech1993} and focused on hyperfine-induced nuclear transitions \cite{PhysRevC.57.3085,PhysRevC.64.064301,PhysRevC.94.014323,PhysRevLett.128.043001,PhysRevResearch.5.023134,Zhang2024NHM}, in which hyperfine mixing modifies the nuclear radiative transition amplitude and thereby alters the isomeric lifetime.
With the development of a general theory of hyperfine-induced nuclear transitions, the mechanism was extended beyond \(^{229}\mathrm{Th}\) to the \(2329~\mathrm{eV}\) isomer of \(^{205}\mathrm{Pb}\), covering highly charged  \(^{205}\mathrm{Pb}^{77+,76+,75+,74+,73+}\) ions 
\cite{Wang2024,Wang2025}. More recently, the studies of NHM have been extended from radiative transitions to NEIES and Coulomb excitation induced by carbon ions ~\cite{Wang2026BeamNHM}, revealing enhancements of several orders of magnitude in the nuclear excitation cross sections of hydrogenlike \(^{229}\mathrm{Th}\) ions compared with those for the bare nucleus.

Positron scattering is a well-established tool in atomic and molecular collision physics, with applications to elastic and inelastic scattering, ionization, positronium formation, and annihilation processes
\cite{Surko2005,RevModPhys.82.2557,Nahar2020,Cheong2024,Mori2024,Machacek2024NeAr}.
Compared with electron scattering, positron scattering provides complementary information on atomic structure and collision dynamics owing to the opposite projectile charge and the absence of exchange with target electrons. Recent advances in positron-beam generation and manipulation~\cite{Danielson2015,Streeter2024,Joshi2025}, including the development of laser-plasma-based sources, have further broadened the experimental capabilities of positron beams. These developments motivate the investigation of positron scattering involving the \(^{229}\mathrm{Th}\) nucleus.

In this work, we develop a relativistic distorted-wave theory for nuclear excitation by inelastic positron scattering (NEIPS) in hydrogenlike \(^{229}\mathrm{Th}^{89+}\) ions, with the NHM effect explicitly included. The target states are described by dressed hyperfine states  \cite{Wang2025,Wang2024-2}, while the incident and outgoing positron states are treated relativistically in the static potential of the target ion. We derive the transition amplitudes associated with the positron-nucleus and positron-electron interactions and calculate the hyperfine-resolved excitation cross sections for incident positron energies ranging from $10^{2}$ to $10^{5}~\mathrm{eV}$. For comparison, NEIPS of the bare \(^{229}\mathrm{Th}^{90+}\) nucleus is also considered to assess the influence of the NHM effect. The present formulation provides a framework for studying NEIPS in other nuclear systems, including highly charged \(^{205}\mathrm{Pb}\) ions.

\section{Theory}\label{sec:Theory}
We consider NEIPS in hydrogenlike $^{229}\mathrm{Th}^{89+}$ ions. The target ion contains one bound
$1s_{1/2}$ electron, whose hyperfine interaction with the nucleus mixes hyperfine states associated with the nuclear ground and isomeric states.
During the collision, the incident positron transfers part of its
kinetic energy to the nucleus, thereby exciting it to the isomeric state.
Atomic units are used throughout this work, namely $\hbar=m_e=|e|=1$.

From Fermi's golden
rule, the differential cross section
for NEIPS is written as
\begin{equation}
\frac{d\sigma_{\mathrm{e}^+}}{d\Omega}
=
\frac{\varepsilon_f\varepsilon_i}{4\pi^2c^4}
\frac{p_f}{p_i}
\left|
\left\langle f\left|
H_{\mathrm{e^+N}}+H_{\mathrm{e^+e^-}}
\right|i\right\rangle
\right|^2 .
\label{eq:dsigma_pos}
\end{equation}
Here, $d\Omega$ is the solid-angle element of the outgoing positron. The quantities 
$p_i$ and $p_f$ are the magnitudes of the incident and outgoing positron momenta, respectively, while $\varepsilon_i$ and
$\varepsilon_f$ are the corresponding total relativistic energies,
$\varepsilon_{i/f}=\sqrt{p_{i/f}^{\,2}c^2+c^4}$.
The kinetic energies of the incident and outgoing positron are denoted by $E_i$ and $E_f$, respectively, and satisfy
$\varepsilon_{i/f}=E_{i/f}+c^2$.
The Hamiltonian $H_{\mathrm{e^+N}}$ describes the electromagnetic interaction
between the projectile positron and the nucleus, while $H_{\mathrm{e^+e^-}}$
describes that between the
positron and the bound electron of the target ion. Here, $\mathrm{e}^+$, $\mathrm{N}$, and $\mathrm{e}^-$ denote the projectile positron, the nucleus, and the bound target electron, respectively.

For the low-energy nuclear transition considered here, the
positron--nucleus interaction can be expanded in electromagnetic
multipoles as
\begin{equation}
H_{\mathrm{e^+N}}
=
\sum_{\tau=E,M}
\sum_{Lm}
(-1)^m
M^{\tau}_{L,-m}
\mathcal{T}^{\tau}_{Lm}.
\label{eq:H_pN}
\end{equation}
Here, the indices $\tau=E$ and $\tau=M$ denote the electric and magnetic
multipole contributions, respectively. The operator
$M^{\tau}_{L,-m}$ is the nuclear multipole operator of rank $L$ and
component $-m$, while $\mathcal{T}^{\tau}_{Lm}$ is the
corresponding projectile multipole operator acting on the positron. Its explicit form is the same as that for an electron \cite{Schwartz1955}, apart from the opposite sign of the projectile charge.
The integer $L$ is
the multipole rank, and $m=-L,\ldots,L$ is its magnetic component.

The interaction between the positron
and the electron is written as
\begin{equation}
\begin{aligned}
&H_{\mathrm{e^+e^-}}\\
=&  \sum_{Lm}
\Bigg[\int d^3r_1\,d^3r_2\,
\rho_{\mathrm{e^+}}(\mathbf r_1)\rho_{\mathrm{e^-}}(\mathbf r_2)C_{Lm}(\hat{\mathbf r}_1)
C^{*}_{Lm}(\hat{\mathbf r}_2)
\frac{r_<^L}{r_>^{L+1}}
\\&
+\frac{1}{c^2L(L+1)}
\int d^3r_1\,d^3r_2\,
\left[\mathbf J_{\mathrm{e^+}}(\mathbf r_1)
\cdot \mathbf L_1C_{Lm}(\hat{\mathbf r}_1)\right]
\\&
\times\left[
\mathbf J_{\mathrm{e^-}}(\mathbf r_2)\cdot
\mathbf L_2C^{*}_{Lm}(\hat{\mathbf r}_2)
\right]
\frac{r_<^L}{r_>^{L+1}}
\Bigg].
\end{aligned}
\label{eq:H_pe}
\end{equation}
Here, $\mathbf r_1$ denotes the coordinate of the projectile positron,
whereas $\mathbf r_2$ denotes that of the bound target electron. We define
$r_i=|\mathbf r_i|$ and $\hat{\mathbf r}_i=\mathbf r_i/r_i$ for
$i=1,2$. The quantities $\rho_{\mathrm{e^+}}(\mathbf r_1)$ and
$\mathbf J_{\mathrm{e^+}}(\mathbf r_1)$ are the charge and current density operators of the projectile positron, respectively, whereas
$\rho_{\mathrm{e^-}}(\mathbf r_2)$ and $\mathbf J_{\mathrm{e^-}}(\mathbf r_2)$ are those of
the bound target electron. The opposite charge signs of the positron and electron are
included implicitly in the definitions of these density
operators.
The operators $\mathbf L_1$ and $\mathbf L_2$ are the orbital angular
momentum operators acting on the coordinates $\mathbf r_1$ and
$\mathbf r_2$, respectively. The function $C_{Lm}(\hat{\mathbf r})$ is
defined by
$C_{Lm}(\hat{\mathbf r})=\sqrt{4\pi/(2L+1)}\,Y_{Lm}(\hat{\mathbf r})$.
Furthermore, we define $r_<=\min(r_1,r_2)$ and 
$r_>=\max(r_1,r_2)$.

The initial and final states are taken as product states
\begin{equation}
\begin{aligned}
|i\rangle&=|F_gM_g\rangle
\left|\chi^{(+)}_{{\mathbf p}_i\eta_i,\mathrm{e}^+}\right\rangle ,\\
|f\rangle&=|F_eM_e\rangle\left|\chi^{(-)}_{{\mathbf p}_f\eta_f,\mathrm{e}^+}\right\rangle,
\end{aligned}
\label{eq:initial_state}
\end{equation}
where $|F_gM_g\rangle$ and $|F_eM_e\rangle$ denote the hyperfine
states associated with the nuclear ground and isomeric states, respectively,
including NHM. The quantum numbers
$F_g$ and $F_e$ are the total angular momenta of the corresponding
hyperfine states, while $M_g$ and $M_e$ are their magnetic projections.
The state $\left|\chi^{(+)}_{{\mathbf p}_i\eta_i,\mathrm{e}^+}\right\rangle$ describes the incident positron with momentum $p_i$ and
spin projection $\eta_i$, whereas
$\left|\chi^{(-)}_{{\mathbf p}_f\eta_f,\mathrm{e}^+}\right\rangle$ describes the scattered positron with momentum $p_f$ and spin projection $\eta_f$. The superscripts ``$+$'' and ``$-$'' specify outgoing- and incoming-wave boundary conditions, respectively.

For hydrogenlike $^{229}\mathrm{Th}^{89+}$, the hyperfine states are described by the dressed hyperfine states \cite{Wang2025,Wang2024-2}
\begin{equation}
\begin{aligned}
|F_gM_g\rangle
=&|I_g\gamma_gJ_g;F_gM_g\rangle
+b_g|I_e\gamma_gJ_g;F_gM_g\rangle,\\
|F_eM_e\rangle
=&|I_e\gamma_gJ_g;F_eM_e\rangle
+b_e|I_g\gamma_gJ_g;F_eM_e\rangle .
\end{aligned}
\label{eq:NHM_states}
\end{equation}
Here, $I_g=5/2$ and $I_e=3/2$ are the spins of the nuclear ground
and isomeric states, respectively. The hyperfine-coupled basis state
$|I_\mu\gamma_gJ_g;F_\mu M_\mu\rangle$ is formed by coupling the nuclear
angular momentum $I_\mu$ to the bound-electron angular momentum $J_g$,
yielding the total angular momentum $F_\mu$ and its projection $M_\mu$. 
The quantum number $\gamma_g$ denotes the remaining quantum numbers
of the bound-electron state. For hydrogenlike
$^{229}\mathrm{Th}^{89+}$, the electron occupies the $1s_{1/2}$ state, so that $J_g=1/2$. Consequently, the nuclear ground-state manifold has $F_g=2,3$, whereas the isomeric-state manifold has $F_e=1,2$. Nuclear hyperfine mixing occurs only between the hyperfine-coupled basis states with $F_g=F_e=2$. The coefficients $b_g$ and $b_e$ characterize the corresponding mixing amplitudes.

In the present
calculation, we use the mixing coefficients reported in Ref.~\cite{Wang2026BeamNHM},
\begin{equation}
\begin{aligned}
F_g&=2, & b_g&=0.048,\\
F_g&=3, & b_g&=0,\\
F_e&=1, & b_e&=0,\\
F_e&=2, & b_e&=-0.048.
\end{aligned}
\label{eq:mixing_coefficients}
\end{equation}
The mixing coefficients are nonzero only for the $F_g=F_e=2$ hyperfine
states. For all other hyperfine states, the mixing is forbidden by angular-momentum coupling, and the
corresponding coefficients therefore vanish.

Within the relativistic distorted-wave approximation, the continuum wave function of the projectile positron is obtained by solving the Dirac equation in the central distorting potential. For a positron with momentum \(\mathbf p\) and spin projection \(\eta\), we denote this state by
$\left|\chi^{(\pm)}_{\mathbf p\eta,\mathrm{e}^+}\right\rangle$ and expand it as \cite{PhysRevLett.124.242501,PhysRevC.106.044604}
\begin{equation}
\begin{aligned}
\left|\chi^{(\pm)}_{\mathbf p\eta,\mathrm{e}^+}\right\rangle
=&
\frac{4\pi}{p}
\sqrt{\frac{\varepsilon+c^2}{2\varepsilon}}
\sum_{\kappa m}
\mathrm{i}^{\,l}
e^{\pm \mathrm{i}\delta^{(+)}_{\kappa}}
\\
&\times
\langle l\,m-\eta,\tfrac12\eta|jm\rangle
Y^{*}_{l\,m-\eta}(\hat{\mathbf p})
\\
&\times
\frac{1}{r}
\begin{pmatrix}
P^{(\mathrm{e}^+)}_{\kappa}(r)\Omega_{\kappa m}(\hat{\mathbf r})
\\[3pt]
\mathrm{i}Q^{(\mathrm{e}^+)}_{\kappa}(r)\Omega_{-\kappa m}(\hat{\mathbf r})
\end{pmatrix},
\end{aligned}
\label{eq:positron_wave}
\end{equation}
where $\varepsilon=\sqrt{p^2c^2+c^4}$ is the total relativistic energy
of the positron, and $\hat{\mathbf p}=\mathbf p/p$ is the direction of
its momentum. The quantities $P_{\kappa}^{(\mathrm{e}^+)}(r)$,
$Q_{\kappa}^{(\mathrm{e}^+)}(r)$, and $\delta_{\kappa}^{(+)}$ denote, respectively, the large radial component, small radial component, and scattering phase shift of the positron continuum state.
The quantity $\Omega_{\kappa m}(\hat{\mathbf r})$ is the spherical
spinor. The spin projection of the positron is $\eta=\pm 1/2$, while
$m$ is the magnetic quantum number associated with the total angular momentum $j$
of the partial wave. The relativistic angular quantum number $\kappa$
is defined by
\begin{equation}
\kappa =
\begin{cases}
-(j+1/2), & l=j-1/2,\\
 +(j+1/2), & l=j+1/2 ,
\end{cases}
\label{eq:kappa_def}
\end{equation}
where $l$ and $j$ denote the orbital and total angular momenta of the positron partial wave, respectively.

The positron radial functions are obtained by solving the stationary Dirac equation
\begin{equation}
\left[
-\mathrm{i}c\boldsymbol{\alpha}\cdot\boldsymbol{\nabla}
+
\beta c^2
+
V_{\mathrm{e}^+}(r)
\right]\psi_{\mathrm{e}^+}
=
\varepsilon\psi_{\mathrm{e}^+},
\label{eq:Dirac_positron}
\end{equation}
where $\psi_{\mathrm{e}^+}$ is the four-component Dirac wave function of the
positron, and
$\boldsymbol{\alpha}=(\alpha_x,\alpha_y,\alpha_z)$ and $\beta$ are the
Dirac matrices.
The central potential $V_{\mathrm{e}^+}(r)$ is the static distorting
potential experienced by the projectile positron in the field of the
target ion. It consists of the repulsive Coulomb potential generated by the finite-size nucleus and the
attractive electrostatic potential generated by the bound $1s_{1/2}$ electron.

After integrating Eq.~\eqref{eq:dsigma_pos} over the outgoing
positron solid angle, averaging over the initial magnetic substates of
the target ion, summing over the final magnetic substates, and applying the partial-wave expansions of the initial and final positron wave functions given in Eq.~\eqref{eq:positron_wave},
the total excitation cross section for the transition  $F_g\rightarrow F_e$ is given by
\begin{equation}
\sigma_{\mathrm{e}^+}
=\frac{2\pi}{c^4}
\frac{p_f}{p_i}
\frac{E_f+c^2}{p_f^2}
\frac{E_i+c^2}{p_i^2[F_g]}
\sum_{\kappa_f,\kappa_i,L}
\frac{\big|
t_{fi}^{\mathrm{e^+N}}
+t_{fi}^{\mathrm{e^+e^-}}
\big|^2}{[L]},
\label{eq:sigma_pos}
\end{equation}
where $[x]\equiv 2x+1$. The kinetic energies $E_i$ and $E_f$ correspond to the incident and outgoing positron momenta \(p_i\) and \(p_f\), respectively.
The amplitudes $t_{fi}^{\mathrm{e^+N}}$ and
$t_{fi}^{\mathrm{e^+e^-}}$ describe the contributions from the
positron-nucleus and positron-electron interactions, respectively. The subscript $fi$ denotes the transition from the initial state $i$ to the final state $f$. In contrast to electron scattering \cite{Wang2026BeamNHM}, no exchange amplitude is present because the projectile positron is distinguishable from the bound target electron. Consequently, the partial-wave summation entering the total cross section is also simpler than in the corresponding electron-scattering formulation.
    
The amplitudes entering Eq.~\eqref{eq:sigma_pos} are
\begin{equation}
\begin{aligned}
t_{fi}^{\mathrm{e^+N}}
=&
\langle F_e \| M_L^E \| F_g \rangle
\langle \kappa_f \| C_L \| \kappa_i \rangle
R_1^{(\mathrm{e}^+)}\\
&+\langle F_e \| M_L^M \| F_g \rangle
\langle \kappa_f \| C_L \| -\kappa_i \rangle
R_2^{(\mathrm{e}^+)},\\
t_{fi}^{\mathrm{e^+e^-}}
=& - A_L \Big[
\langle \kappa_g \| C_L \| \kappa_g \rangle
\langle \kappa_f \| C_L \| \kappa_i \rangle
I_1^{(\mathrm{e}^+)}\\
&-\langle \kappa_g \| C_L \| -\kappa_g\rangle
\langle \kappa_f \| C_L \| -\kappa_i \rangle
I_2^{(\mathrm{e}^+)}
\Big].
\end{aligned}
\label{eq:t_matrix_pos}
\end{equation}
Here, $\kappa_i$ and $\kappa_f$ are the relativistic angular quantum
numbers of the incident and outgoing positron partial waves,
respectively.  The quantity $\kappa_g$ denotes the relativistic angular
quantum number of the bound target electron. For the $1s_{1/2}$
target electron, $\kappa_g=-1$. The reduced matrix elements
$\langle F_e \| M_L^E \| F_g \rangle$ and
$\langle F_e \| M_L^M \| F_g \rangle$ are the electric and magnetic
nuclear multipole transition matrix elements between the initial and final hyperfine states, respectively. The quantity $\langle \kappa_f \| C_L \| \kappa_i \rangle$ is the reduced matrix element of the rank-\(L\) spherical tensor $C_L$ between the initial and final positron partial waves. 

The radial integrals entering Eq.~\eqref{eq:t_matrix_pos} are defined
as
\begin{equation}
\begin{aligned}
R_1^{(\mathrm{e}^+)}
&=
\int_0^\infty
\frac{f_1^{(\mathrm{e}^+)}(r)}{r^{L+1}}\,dr,
\\
R_2^{(\mathrm{e}^+)}
&=
\frac{\kappa_i+\kappa_f}{L}
\int_0^\infty
\frac{f_2^{(\mathrm{e}^+)}(r)}{r^{L+1}}\,dr,
\\
I_1^{(\mathrm{e}^+)}
&=
\int_0^\infty\!\!\int_0^\infty
f_1^{(\mathrm{e}^+)}(r_1)g_1(r_2)
\frac{r_<^L}{r_>^{L+1}}
\,dr_1\,dr_2,
\\
I_2^{(\mathrm{e}^+)}
&=
\frac{2\kappa_g(\kappa_i+\kappa_f)}{L(L+1)}
\int_0^\infty\!\!\int_0^\infty
f_2^{(\mathrm{e}^+)}(r_1)g_2(r_2)\\
&\times
\frac{r_<^L}{r_>^{L+1}}
\,dr_1\,dr_2.
\end{aligned}
\label{eq:radial_integrals_pos}
\end{equation}
The radial
functions appearing in Eq.~\eqref{eq:radial_integrals_pos} are
\begin{equation}
\begin{aligned}
f_1^{(\mathrm{e}^+)}(r)
&=
P_{\kappa_f}^{(\mathrm{e}^+)}(r)P_{\kappa_i}^{(\mathrm{e}^+)}(r)
+
Q_{\kappa_f}^{(\mathrm{e}^+)}(r)Q_{\kappa_i}^{(\mathrm{e}^+)}(r),
\\
f_2^{(\mathrm{e}^+)}(r)
&=
P_{\kappa_f}^{(\mathrm{e}^+)}(r)Q_{\kappa_i}^{(\mathrm{e}^+)}(r)
+
P_{\kappa_i}^{(\mathrm{e}^+)}(r)Q_{\kappa_f}^{(\mathrm{e}^+)}(r),
\\
g_1(r)
&=
P_{\kappa_g}^{2}(r)
+
Q_{\kappa_g}^{2}(r),
\\
g_2(r)
&=
2P_{\kappa_g}(r)Q_{\kappa_g}(r),
\end{aligned}
\label{eq:radial_functions_pos}
\end{equation}
where \(P_{\kappa_i}^{(\mathrm{e}^+)}(r)\) and \(Q_{\kappa_i}^{(\mathrm{e}^+)}(r)\) are the large and small radial components of the incident positron partial wave, respectively, while \(P_{\kappa_f}^{(\mathrm{e}^+)}(r)\) and \(Q_{\kappa_f}^{(\mathrm{e}^+)}(r)\) are the corresponding components of the outgoing positron partial wave. The functions $P_{\kappa_g}(r)$ and $Q_{\kappa_g}(r)$ are the large and small radial components of the bound \(1s_{1/2}\) electron, respectively.

\begin{figure}
\centering
\includegraphics[width=0.95\linewidth]{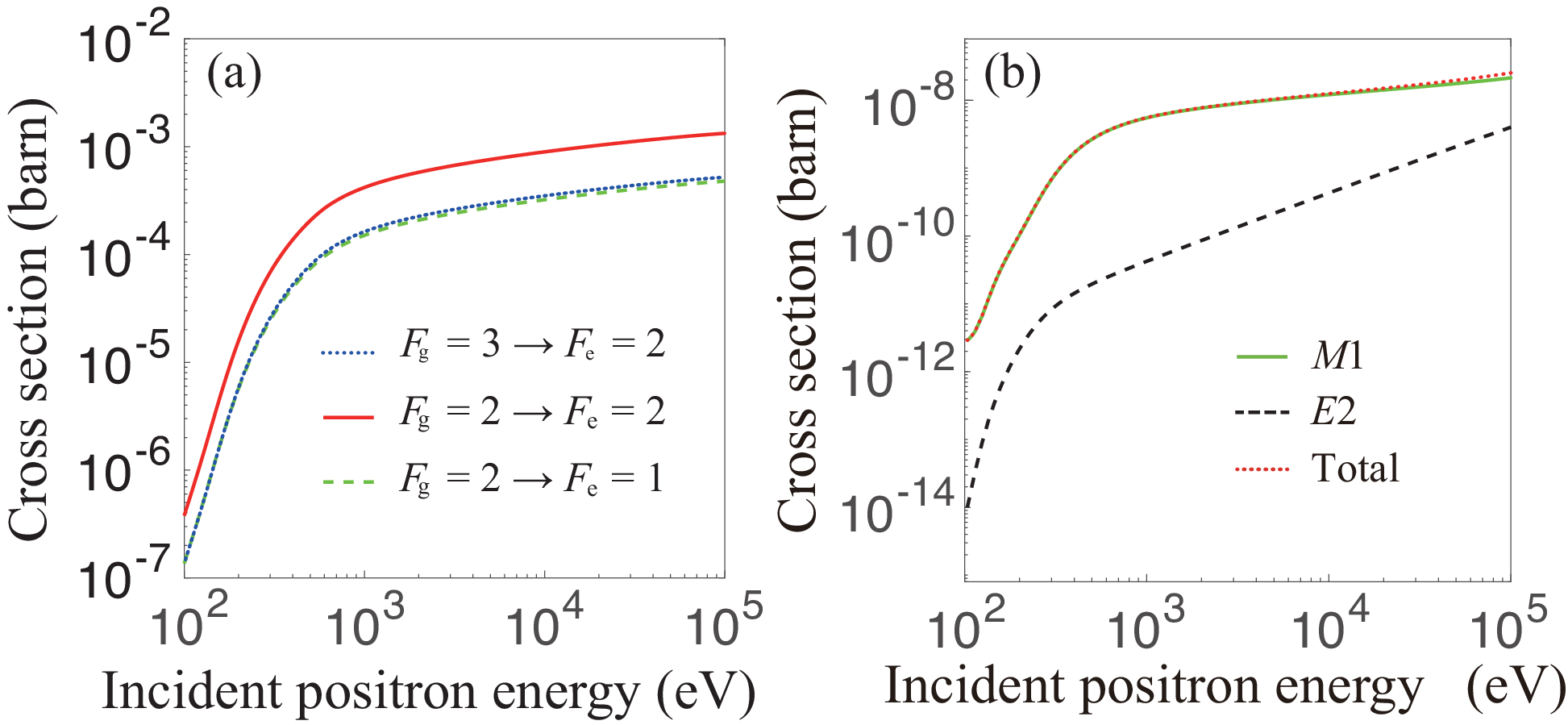}
\caption{Cross sections for NEIPS 
as functions of the incident positron energy from $10^2$ to
$10^5$ eV. 
(a) NEIPS cross sections in hydrogenlike
$^{229}\mathrm{Th}^{89+}$ ions with the NHM effect included. Three
hyperfine excitation channels are shown as labeled. 
(b) NEIPS cross sections for the bare $^{229}\mathrm{Th}^{90+}$
nucleus, where the NHM effect is absent. The total cross section and
the individual $M1$ and $E2$ contributions are shown.}
\label{fig:positron_channels}
\end{figure}

The angular coefficient associated with the dressed hyperfine states
is denoted by $A_L$ and is given by
\begin{equation}
\begin{aligned}
A_L
=&
\sqrt{[F_e][F_g]}
\Bigg[
b_g
(-1)^{J_g+L+F_e+I_e}
\begin{Bmatrix}
J_g& J_g & L \\
F_g & F_e & I_e
\end{Bmatrix}
\\
&\quad 
+b_e^{*}
(-1)^{J_g+L+F_e+I_g}
\begin{Bmatrix}
J_g & J_g & L \\
F_g & F_e & I_g
\end{Bmatrix}
\Bigg].
\end{aligned}
\label{eq:A_L_pos}
\end{equation}
Here, $J_g$ is the total angular momentum associated with the relativistic angular quantum number \(\kappa_g\) and the braces denote Wigner $6j$ symbols.
The coefficient $A_L$
arises from the angular-momentum recoupling between the nuclear and
electronic degrees of freedom in the dressed hyperfine states.
Accordingly, the hyperfine mixing coefficients $b_g$ and
$b_e$ enter explicitly in $A_L$. In the absence of hyperfine
mixing, $b_g=b_e=0$, so that $A_L=0$ and hence
$t_{fi}^{\mathrm{e^+e^-}}=0$. The excitation cross section $\sigma_{\mathrm{e}^+}$ then
reduces to that of the direct positron-nucleus inelastic scattering
process.


\section{results}\label{sec:results}
In the hydrogenlike $^{229}\mathrm{Th}^{89+}$ ion, the bound $1s_{1/2}$ electron splits the nuclear ground state with $I_g^\pi=5/2^+$ into the hyperfine levels $F_g=2,3$, and the isomeric state with $I_e^\pi=3/2^+$ into the hyperfine levels $F_e=1,2$. In the following calculations, the reduced nuclear transition matrix elements are obtained from the reduced transition probabilities $B(M1,I_e\rightarrow I_g)=0.022~\mathrm{W.u.}$~\cite{PhysRevLett.132.182501} and $B(E2,I_e\rightarrow I_g)=30~\mathrm{W.u.}$~\cite{Minkov2021}.

The radial wave functions of the bound electron and projectile positron are obtained using the RADIAL package \cite{Salvat2019} by solving the Dirac equation \eqref{eq:Dirac_positron}. For hydrogenlike \(^{229}\mathrm{Th}^{89+}\), the positron potential consists of the finite-size nuclear Coulomb potential and the electrostatic potential generated by the bound \(1s_{1/2}\) electron. For bare \(^{229}\mathrm{Th}^{90+}\), only the finite-size nuclear Coulomb potential is included. The continuum radial wave functions are normalized according to the asymptotic condition \(P_{\kappa}(r)\propto \sin(pr-l\pi/2+\delta_\kappa^\prime)\) as $r\to\infty$, where $\delta_\kappa^\prime$ denotes the scattering phase shift. 

Once the bound-state and continuum wave functions are obtained, the NEIPS cross sections in hydrogenlike \(^{229}\mathrm{Th}^{89+}\) can be calculated using Eq.~\eqref{eq:sigma_pos}. The results are shown in Fig.~\ref{fig:positron_channels}(a) for three hyperfine excitation channels: \(F_g=2\rightarrow F_e=2\), \(F_g=2\rightarrow F_e=1\), and \(F_g=3\rightarrow F_e=2\). Over the incident positron energy range considered,
the cross sections for all three channels increase monotonically from approximately \(10^{-7}\) to \(10^{-3}\) barn. This behavior is mainly attributed to the repulsive Coulomb interaction between the positron and the highly charged target ion. At low incident energies, the Coulomb barrier suppresses the positron wave function in the inner region of the target, where both the positron-nucleus interaction and the positron-electron interaction through the corresponding charge and current densities are strongest. As the
incident energy increases, this suppression is reduced, resulting in greater overlap of the positron wave function with these inner-region densities and, consequently, a rapid increase in the excitation cross section. At higher energies, the penetration becomes less sensitive to the incident energy, and the increase in the cross section gradually slows.

Among the three channels, the $F_g=2\rightarrow F_e=2$ transition exhibits the largest cross section. Since both mixing coefficients, \(b_g\) for \(F_g=2\) and \(b_e\) for \(F_e=2\), are nonzero, this channel receives the strongest NHM contribution. In contrast, the \(F_g=2\rightarrow F_e=1\) and \(F_g=3\rightarrow F_e=2\) channels involve only one nonzero mixing coefficient, leading to smaller and nearly identical cross sections over the energy range considered. The \(F_g=3\rightarrow F_e=1\) channel is not shown because both mixing coefficients vanish, with \(b_g=0\) for \(F_g=3\) and \(b_e=0\) for \(F_e=1\). Consequently, the NHM contribution is absent, leaving only the much weaker direct positron-nucleus excitation.

To assess the contribution of NHM relative to direct positron-nucleus excitation, we also consider NEIPS in bare \(^{229}\mathrm{Th}^{90+}\), where no bound electron is present and NHM is therefore absent. The corresponding results are shown in Fig.~\ref{fig:positron_channels}(b). In this case, excitation proceeds solely through the direct positron-nucleus interaction. The total cross section is shown together with the individual \(M1\) and \(E2\) contributions. The direct excitation is dominated by the \(M1\) transition, whereas the \(E2\) contribution, though much smaller at low positron energies, increases with energy and becomes non-negligible at the upper end of the energy range. Nevertheless, the total bare-nucleus cross section is several orders of magnitude smaller than that obtained for hydrogenlike $^{229}\mathrm{Th}^{89+}$ when NHM is included. This comparison demonstrates that bound-electron-induced NHM provides an efficient additional pathway for positron-induced isomeric excitation.

\section{conclusion}\label{sec:conclusio}
In conclusion, we have developed a relativistic theory for NEIPS that incorporates the NHM effect, and applied it to hydrogenlike \(^{229}\mathrm{Th}^{89+}\). Based on this framework, we evaluated the hyperfine-resolved excitation cross sections and found that they increase monotonically from approximately \(10^{-7}\) to \(10^{-3}\) barn over the incident positron energy range considered. Among the different channels, the \(F_g=2\rightarrow F_e=2\) transition exhibits the largest cross section. For comparison, NEIPS in bare \(^{229}\mathrm{Th}^{90+}\), where NHM is absent, yields cross sections four orders of magnitude smaller than those in hydrogenlike \(^{229}\mathrm{Th}^{89+}\). These results broaden the range of nuclear processes in which NHM can play a significant role and establish NHM as an effective mechanism for enhancing NEIPS cross sections in highly charged ions.

\begin{acknowledgments}
This work was supported by the Strategic Priority Research Program of the Chinese Academy of Sciences (Grant No.~XDB0920000), the National Natural Science Foundation of China (Grants No.~12574387 and No.~12405026), the Innovation Program for Quantum Science and Technology (Grant No.~2023ZD0300704), and the Funds of the Natural Science Foundation of Hangzhou (Grant No.~2024SZRYBA050001).
\end{acknowledgments}



\bibliographystyle{apsrev4-2}
\bibliography{references}


\end{document}